\documentclass[review,jog]{igs}

\usepackage[utf8]{inputenc}
\usepackage[hidelinks]{hyperref}
\usepackage{amssymb}
\usepackage{graphicx}
\usepackage{siunitx}
\usepackage[capitalise]{cleveref}
\usepackage[normalem]{ulem}

\usepackage{placeins}

\jourvolume{0}
\jourissue{0}
\jourpubyear{2025}

\begin{document}

\title{Modeling ice cliff stability using a new Mohr-Coulomb-based phase field fracture model}

\author[Clayton et. al.]{Theo CLAYTON,$^1$
   Ravindra DUDDU,$^{2}$ Tim HAGEMAN,$^3$  
  Emilio MART\'{I}NEZ-PA\~NEDA $^{3,1}$}

\affiliation{%
$^1$Department of Civil and Environmental Engineering, Imperial College London, London SW7 2AZ, UK\\
$^2$Department of Civil and Environmental Engineering, Department of Earth and Environmental Sciences, Vanderbilt University, Nashville, TN 37235, USA\\
$^3$ Department of Engineering Science, University of Oxford, Oxford OX1 3PJ, UK \\
  Correspondence: Ravindra Duddu \email{ravindra.duddu@vanderbilt.edu}, \\Emilio Mart\'{\i}nez-Pa\~neda 
  \email{emilio.martinez-paneda@eng.ox.ac.uk}}

\begin{frontmatter}

\maketitle

\begin{abstract}
Iceberg calving at glacier termini results in mass loss from ice sheets, but the associated fracture mechanics is often poorly represented using simplistic (empirical or elementary mechanics-based) failure criteria. Here, we propose an advanced Mohr-Coulomb failure criterion that drives cracking based on the visco-elastic stress state in ice. This criterion is implemented in a phase field fracture framework, and finite element simulations are conducted to determine the critical conditions that can trigger ice cliff collapse. Results demonstrate that fast-moving glaciers with negligible basal friction are prone to tensile failure causing crevasse propagation far away from the ice front; whilst slow-moving glaciers with significant basal friction are likely to exhibit shear failure near the ice front. Results also indicate that seawater pressure plays a major role in modulating cliff failure. For land terminating glaciers, full thickness cliff failure is observed if the glacier exceeds a critical height, dependent on cohesive strength $\tau_\mathrm{c}$ ($H \approx 120\;\text{m}$ for $\tau_\mathrm{c}=0.5\;\text{MPa}$). For marine-terminating glaciers, ice cliff failure occurs if a critical glacier free-board ($H-h_\mathrm{w}$) is exceeded, with ice slumping only observed above the ocean-water height; for $\tau_\mathrm{c} = 0.5\;\text{MPa}$, the model-predicted critical free-board is $H-h_\mathrm{w} \approx 215\;\text{m}$, which is in good agreement with field observations. While the critical free-board height is larger than that predicted by some previous models, we cannot conclude that marine ice cliff instability is less likely because we do not include other failure processes such as hydrofracture of basal crevasses and plastic necking.
\end{abstract}

\end{frontmatter}

\section{Introduction} \label{introduction}
Glacial mass loss from the Greenland and Antarctic ice sheets has become a significant contributor to sea level rise, with the main processes of mass loss being iceberg calving events and ocean-induced melting at the underside of ice shelves \citep{Frederikse2020, Siegert2020}. The processes of ablation have accelerated over the last decades due to atmospheric and oceanic warming and have exceeded ice accumulation rates via snowfall, leading to net mass loss from Greenland and Antarctic ice sheets \citep{Benn2010, Rignot2019}. However, describing the mass loss due to fracture-induced iceberg calving events is complex as it is coupled with the viscoelastic deformation process, requiring advanced physics-based models for crevasse nucleation and propagation under mixed mode loading, and taking into account the effects of meltwater on this propagation.

Atmospheric warming enhanced by anthropogenic carbon emissions has led to increased production of surface meltwater \citep{Mercer1978}, stored in supraglacial lakes and firn aquifers \citep{Poinar2017}. Meltwater may accumulate in surface crevasses and apply additional tensile stresses to crevasse walls \citep{Weertman1971,VanderVeen2007}, and cause hydrofracture in Greenland glaciers \citep{Das2008,Hageman2024}. This hydrofracture process can also lead to increased crevasse propagation in ice shelves and has the potential to cause large iceberg calving events \citep{Scambos2009} and affects the future stability of Antarctic ice shelves and glaciers \citep{bassis2024stability}. 

In the regions of Western Antarctica such as Pine Island Glacier and Thwaites Glacier, where ice is grounded well below sea level and the glacier sits on a retrograde bed slope, where the bed deepens upstream. As a result, if ice sheet regression occurs beyond a critical ``tipping point'', irreversible rapid grounding line retreat is likely to occur \citep{Weertman1974b, Hill2023}: As the ice progressively gets thicker inland, and ice flux is a function of ice thickness, the rate of ice loss increases as the grounding line retreats \citep{Schoof2007}. This process is termed as the marine ice sheet instability (MISI). Distinctively, the removal of floating ice shelves would expose ice cliffs at the grounding line, which if sufficiently tall, are prone to structural failure. Similarly to MISI, if the grounded ice sheet is located on a retrograde bed slope, progressively thicker cliff faces will become exposed, leading to a potential rapid grounding line retreat; a theory known as the marine ice cliff instability (MICI). However, both MICI and MISI to some extent remain controversial theories and are yet to be directly observed \citep{Wise2017, Edwards2019}.

The conditions under which these fracture events occur are different: Antarctic glaciers tend to have floating ice-shelves at the terminus, protecting their grounded region from full-thickness crevasses through buttressing effects. In contrast, the termini of ocean-terminating Greenland glaciers are typically grounded, producing steep ice-cliffs with only limited support from the oceanwater pressure, making full-thickness fracture in the grounded regions of the ice sheet more likely. While the same rheological and fracture/calving models are applicable to glaciers in both Greenland and Antarctic ice sheets, the boundary conditions and the associated failure mechanics can be different. Here, we will limit the application of our developed models to grounded glaciers, which is representative of the conditions in Greenland.

The majority of analytical and numerical fracture analyses in glacial ice have considered fracture to be purely tensile, with crevasses propagating vertically downward as a result of the far-field longitudinal stress state. These include the classical analytical methods based on Nye zero stress  \citep{Nye1957} and linear elastic fracture mechanics \citep{VanDerVeen1998}, as well as numerical methods based on continuum damage mechanics \citep{Duddu2020,Huth2021b}, cohesive zone \citep{gao2023finite,Hageman2024} and phase field fracture \citep{Sun2021,Clayton2022,nguyen2025adaptive}. However, a mode of failure that has been largely unexplored is that of mixed mode I-II fracture under shear and compressive loading \citep{Schulson2001, Schulson1999}. As this shear failure leading to mode II (sliding) fracture is the mechanism upon which MICI is based \citep{Bassis2012}, it becomes important to extend traditional mode I (opening) fracture models to capture both tension and shear failure. 

\citet{Bassis2012} considered the combination of shear and tensile failures to determine a semi-empirical upper and lower bound for maximum cliff height, depending on the presence of crevasses. For a land terminating cliff without initial crevasses or meltwater, the maximum cliff height was calculated as $H_{\text{max}} = 220 \; \text{m}$ for a depth-averaged yield strength of $1 \; \text{MPa}$, this reduces to $100 \; \text{m}$ with the presence of crevasses. \citet{Parizek2019} argued that \textit{the threshold cliff height for slumping is likely to be slightly above 100 m in many cases, and roughly twice that (145–285 m) in mechanically competent ice under well-drained or low-melt conditions}. In contrast, \citet{Clerc2019} argued that as the ice-shelf removal timescale increases, viscous relaxation dominates, and the critical height increases to ~540 m for timescales greater than days; whereas the 90-m critical height implies ice-shelf removal in under an hour. 

The empirical calving law suggested by \citet{Bassis2012} has been coupled with ice sheet models to find that MICI has the potential to accelerate the collapse of the Western Antarctic Ice Sheet to a time-scale of decades, resulting in a global sea level rise of $15 \; \text{m}$ within a few hundred years \citep{Pollard2015, DeConto2016}. However, these findings have recently been contrasted by the results from \citet{Morlighem2024}, who demonstrated that the inclusion of the same ice-cliff stability criterion did not result in rapid ice sheet retreat for the Twaites Glacier using several ice sheet models. The role of glacial bed conditions has been investigated by \citet{Ma2017} and \citet{Benn2017}, solving for the full Stokes equations in 2D  using finite elements, and finding that glaciers subject to free slip are dominated by tensile failure and no-slip glaciers are subject to shear failure. In addition, calving laws have been suggested based on the maximum shear stress in no-slip glaciers, to determine the maximum free-board of ice cliffs that may be sustained \citep{Schlemm2019}.  

To simulate these fracture processes, the discrete element method (DEM) has been used to simulate the brittle failure of ice cliffs via shear failure \citep{Benn2017, Crawford2021, Bassis2021}. These methods consider solids as a mass of discrete particles, with forces being transmitted through the particles via elastic bonding \citep{Wang2006}. Particle bonding may be represented through elastic beams and failure occurs when bonds break under the Mohr-Coulomb failure criterion \citep{Astrom2013}. The DEM has been used to capture the complex fracture patterns occurring during ice-cliff collapse events, capturing both the alternating surface and basal cracking that MICI is based on \citep{Benn2018}. However, the DEM is computationally expensive and linking the particle interactions to physical parameters is complicated, hindering the incorporation of the non-linear viscous creep deformation of ice. 

In contrast with the DEM, continuum damage mechanics methods, including the phase field method, avoid the representation of each and every smaller-scale crack and instead capture the larger-scale fracture morphology, thereby leading to a smeared crack description. Thus, these methods are able to limit the computational cost of fracture simulations whilst allowing the use of viscoelastic and full Stokes formulations \citep{Jimenez2017}. Notably, the phase field fracture method has recently been used to study the propagation of crevasses in glaciers and ice shelves \citep{Sun2021, Clayton2022, Sondershaus2022}.  
Initially developed for brittle fracture in elastic media \citep{Miehe2010}, phase field methods may be coupled with other physical processes to solve complex multi-physics problems including hydraulic fracture \citep{Zhou2018,Zhou2019}, corrosion damage \citep{gao2020space,Kovacevic2023} and hydrogen embrittlement \citep{martinez2018phase,wu2020phase} among others. 
As the location of cracks does not need to be known beforehand, the phase field fracture method allows us to investigate both surface and basal crevasse (damage) nucleation, propagation and coalescence in glaciers in relation to the geometry and boundary conditions, which play a key role in determining calving patterns \citep{duddu2013numerical,bassis2013diverse}. 

In this paper, we extend the phase field fracture method to capture ice-cliff instabilities by introducing a new stress-based crack driving force function based on the Mohr-Coulomb failure criterion. This driving force will allow shear, tensile, and mixed-mode fractures to be captured, thereby representing all the mechanisms that potentially drive ice-cliff instabilities. We then perform finite element simulations to assess the structural failure of ice cliffs at the terminus of grounded glaciers. By considering variations in basal slip boundary conditions, glacier thickness $H$, ocean-water height $h_{\text{w}}$ and ice strength parameters, the criteria in which ice cliffs become unstable are determined. Finally, we compare these results to empirical relations derived in the literature and discuss how the stability criteria resulting from simulations can be used to inform ice sheet modellers on what conditions are required to trigger ice cliff failure.   

\section{Constitutive Theory}

\begin{figure}
 \centering
  \includegraphics[width=0.48\textwidth]{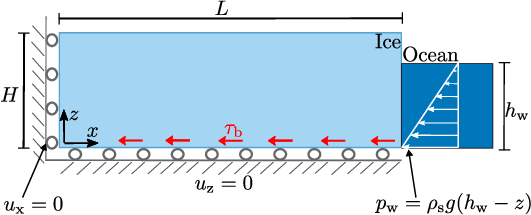}
  \caption{Schematic diagram showing the coordinate system and dimensions used for a grounded glacier, and the boundary conditions subject to the following basal conditions: free slip ($\tau_\text{b}=0)$, basal friction ($\tau_\text{b}$ following \cref{eqn:Weertman_Sliding}, or a frozen/fixed basal boundary condition ($u_x=0$ on the bottom boundary).}
\label{fig:Boundary_Conditions_MICI}
\end{figure}

In this section, we present the constitutive theory for solving the coupled fracture-deformation problem for simulating ice cliff failure. The two key components of this formulation are: (1) elastic and non-linear viscous constitutive models, and (2) a phase field fracture model driven by shear stresses in addition to the hydrostatic pressure. The independent variables solved for are the displacement $\mathbf{u}$ and phase field variable $\phi$. We assume that ice is under isothermal conditions, such that thermo-mechanical melting/refreezing is neglected. We also ignore the role of meltwater in driving crevasses by assuming dry crevasses, such that predictions on the stability of ice cliffs pose an upper limit on their actual stability. Finally, we employ the plane strain assumption, allowing the three-dimensional geometry to be reduced to a two-dimensional geometry along a flow line with the $x$ and $z$ denoting the horizontal (along flow) and vertical coordinates (see \cref{fig:Boundary_Conditions_MICI}). This is a reasonable assumption for some glaciers where the out-of-plane dimension is typically larger than the in-plane length, with the majority of strains occurring in this $(x,z)$ plane.

Assuming small strains and rotations, the total strain tensor is given by the symmetric gradient of the displacement field as:
\begin{equation}
    \mathbf{\varepsilon} = \frac{1}{2}\left(\nabla\mathbf{u}^T+\nabla\mathbf{u}\right)  ,
\end{equation}
Ice rheology may be characterised through a Maxwell visco-elastic model, 
wherein the total strain is additively decomposed into elastic ($\mathbf{\varepsilon}^\text{e}$) and viscous ($\mathbf{\varepsilon}^\text{v}$) components:
\begin{equation}
    \mathbf{\varepsilon} = \mathbf{\varepsilon}^\text{e} + \mathbf{\varepsilon}^\text{v} \, .
\end{equation}
Over shorter timescales, linear elastic deformation is the governing mode of deformation \citep{Christmann2016}, whereas over long timescales and in the presence of stress concentrations \citep{Hageman2024}, glacial ice is known to behave as a non-linear viscous fluid, as it is a polycrystalline material operating close to its melting point. 

The viscous strain rate $\Dot{ \boldsymbol{\varepsilon}}^\text{v}$ is calculated using the Glen's flow law \citep{Glen1955}:
\begin{equation} \label{eq:GlenLaw}
    \Dot{ \mathbf{\varepsilon}}^\text{v} = A \left( \sigma_\text{e}\right)^{n-1} \mathbf{\sigma}_0' 
\end{equation}
where $A$ is the creep coefficient, $n$ is the creep exponent, $\mathbf{\sigma}_0'$ is the undamaged deviatoric stress tensor, $\sigma_\text{e}$ is the equivalent stress, equal to the second invariant of the deviatoric stress  $\sigma_\text{e} = \sqrt{\frac{1}{2}\mathbf{\sigma}_0' : \mathbf{\sigma}_0'}$. The creep coefficient $A$ and creep exponent $n$ are mechanical properties found through experimental data, with $A$ exhibiting a temperature dependency through the Arrhenius law 
\begin{equation}
    A = A_{0} \exp{\left(\frac{Q}{R}\left( \frac{1}{T} - \frac{1}{T_{0}}\right) \right)} \, ,
\end{equation}
where $T$ is the absolute temperature, $Q$ is the activation energy, $R$ is the universal gas constant, and $A_{0}$ is the creep coefficient at a reference temperature $T_{0}$.

Throughout this work, we assume the material properties of ice to be constant, independent of the depth. While it has been shown that including this depth-dependence plays an important role \citep{Clayton2024}, we chose to focus solely on the role of the ice-sheet thickness and basal conditions on the propagation of shear fractures. However, the presented model can be easily adapted to include depth-dependent material parameters.

\subsection{Phase Field Theory}

Phase field fracture is used to describe the damage process. The phase field fracture method has received significant attention in recent years, as it provides a computationally compelling and physically sound approach of predicting the evolution of cracks, based on Griffith's energy balance and the thermodynamics of fracture \citep{griffith1921vi,Bourdin2000}. Phase field methods overcome the challenges associated with computationally tracking evolving interfaces by smearing interfaces over a finite domain, defined through a phase field length scale $\ell$, and describing their evolution through an auxiliary phase field variable $\phi$. In the case of fracture problems, the phase field variable takes a value of $\phi=0$ in the intact state and a value of $\phi=1$ in fully cracked material points, varying smoothly in-between these two phases. Various phase field fracture models have been proposed through the years. Of particular relevance here are the so-called stress-based phase field fracture models, which are less sensitive to the length scale magnitude \citep{Miehe2015a}, enabling the use of coarser meshes, as required to simulate large ice sheet domains \citep{Clayton2024}. In stress-based models, the fracture driving force is typically given as a function of the principal stress as, 
\begin{equation}\label{eq:Dd}
    D_\mathrm{d} = \zeta \left\langle \sum_{\textrm{a}=1}^3 \left( \frac{\langle \tilde{\sigma}_\textrm{a} \rangle}{\sigma_\textrm{c}} \right)^2 - 1 \right \rangle
\end{equation}
where $\tilde{\sigma}_\textrm{a}$ is the principal stress in each direction, $\zeta$ is a parameter governing the behaviour in the post-failure region, and $\sigma_\textrm{c}=\sqrt{3G_\text{c}E/8\ell}$ is the material strength \citep{kristensen2021assessment}. The Macaulay brackets $\langle\square\rangle$ are used to indicate that only positive (extensional) stresses contribute to crack propagation. To ensure damage irreversibility, a history field $H_\textrm{d}$ is defined such that,
\begin{equation}
    H_\textrm{d} = \max_{\tau \in[0,t]} D_\textrm{d}
\end{equation}
This history field ensures that the driving force for crack propagation is either constant or increasing over time, thus ensuring that crevasses do not heal if unloading occurs. Note that a reduction in driving force would violates the thermodynamic constraint of damage irreversibility, unless the healing due to refreezing or viscous flow.

Then, the evolution equation for the phase field is expressed as follows \citep{Miehe2015a},
\begin{equation}\label{eq:PhiStrongForm}
   2 \left( 1 - \phi \right) H_\textrm{d} - \left( \phi - \ell^2 \nabla^2 \phi \right) = \eta \Dot{\phi} 
\end{equation}
A rate-dependent term is introduced into the phase field evolution law \cref{eq:PhiStrongForm}, which contains the phase field viscosity term $\eta$ \citep{Miehe2010}. Introducing rate dependency in gravitational-driven fracture problems improves numerical convergence as it prevents excessive damage evolving due to the instantaneous stress field. In previous work, rate dependency was introduced by including inertial terms in the momentum balance \citep{Clayton2022} with a similar effect on damage evolution.

The failure surface for the crack driving force based on principal stresses in \cref{eq:Dd} is plotted in \cref{fig:Yield_Surface} for $\zeta = 1$. If the stress state lies within the shade region of the graph, the material will remain intact. In contrast, if the stress is at the boundary of the surface, any further loading will develop fractures within the ice. The presence of the Macaulay brackets in \cref{eq:Dd} results in a tension-only type failure in regions where only either $\sigma_{xx}$ or $\sigma_{zz}$ is tensile, with yielding occurring above the fracture stress $\sigma_\textrm{c}$. When both $\sigma_{xx}$ and $\sigma_{zz}$ are tensile, the failure surface is bounded by a quadratic barrier function, the size of which is dependent on $\sigma_\mathrm{c}$, and the material's behaviour during the fracture process being dependent on $\zeta$. Recently, \citet{gupta2024damage} demonstrated the ability of the principal stress-based driving force to accurately capture fracture propagation in 3D printed rock samples.

The phase field evolution law is coupled with the momentum balance:
\begin{equation}\label{eq:DispStrongForm}
    \nabla \cdot \left\{ (1-\phi)^2 \mathbf{C}_0 \left( \mathbf{\varepsilon} - \mathbf{\varepsilon}^v \right) \right\} + (1-\phi)^2\rho_\text{i}\mathbf{g}= 0
\end{equation}
where $\mathbf{C}_0$ denotes the linear-elastic stiffness matrix of undamaged ice, and the self-weight of ice is characterised by $\rho_\text{i}\mathbf{g}$. The phase-field variable $\phi$ degrades both the stiffness and the body force term. As a result when the glacier fractures, the damaged ice no longer transfers any loads. As we also degrade the gravity term (to prevent large free-body motion when the ice is fully damaged), this effectively removes the damaged parts from impacting the still intact parts of the glacier.

The model can readily be extended to incorporate the effect of hydrostatic pressure in water-filled crevasses, through a poro-damage approach \citep{Mobasher2016,Sun2021}. This effect is neglected here, as it is difficult to constrain the amount of water in crevasses near ice cliffs based on observations. As a result, the stability limits obtained in our results section will approximate an upper limit on the stability, with the addition of meltwater always reducing the maximum cliff heights compared to our results obtained for dry crevasses. 

Together, \cref{eq:DispStrongForm,eq:PhiStrongForm} define the governing equations of the fracture--deformation problem. These equations are solved in COMSOL Multiphysics using a multi-pass staggered solver, for each time increment iterating between solving for displacements and phase-field. Convergence of the dependent variables is achieved when the error between iterations is less than the relative tolerance $r = 0.001$ for both variables before proceeding to the next time increment. Quadratic triangular mesh elements are used to discretise the domain, and an adaptive time-stepping scheme is utilised. 

\subsection{Mohr-Coulomb Based Crack Driving Force}

While \cref{eq:Dd} is a commonly used crack driving force, it is only able to capture mode-I (tensile) cracks. Herein, we propose an alternative crack driving force based on a Mohr-Coulomb failure criterion, to describe brittle compressive failure in response to shear stresses, inspired by the work of \cite{Schlemm2019}. In the general form, shear stress $\tau_\text{n}$ acting on a plane with normal $\mathbf{n}$ are resisted by a combination of the material's cohesive strength $\tau_{\mathrm{c}}$ and internal friction $\mu$ acting based on the normal stress $\sigma_\text{n}$, such that the fracture criterion is given as: 
\begin{equation}
f_\text{c} = |\tau_\text{n}| - \mu \sigma_\text{n} -\tau_{\textrm{c}}
\end{equation}
with fracture occurring when $f_\text{c}>0$.

This Mohr-Coulomb failure criterion may be rewritten in terms of maximum shear stress $\tau_{\mathrm{max}}=|\tau_\text{n}|/\sqrt{1+\mu^2}$ and pressure $P$ as \citep{Schulson2001, Jaeger1979}:
\begin{equation}
f_\text{c} = \sqrt{\mu^2 + 1} \; \tau_{\textrm{max}} - \mu P -\tau_{\textrm{c}}
\end{equation}
In the 2D plane strain case, the  maximum shear stress $\tau_{\textrm{max}}$ is given by 
\begin{equation} \label{eq:tau_max}
\tau_{\textrm{max}} = \sqrt{\left( \frac{\sigma_{xx}-\sigma_{zz}}{2}\right)^2 + \sigma_{xz}^2}
\end{equation}
and is the equivalent to the radius of the Mohr circle.  This operates at $45 ^\circ$ to the maximum principal stress $\sigma_{1}$
\begin{equation} \label{eq:sigma_1}
\sigma_{1} = \frac{\sigma_{xx}+\sigma_{zz}}{2}+\sqrt{\left(\frac{\sigma_{xx}-\sigma_{zz}}{2} \right)^2+\sigma_{xz}^2}
\end{equation}
The isotropic pressure $P$ is given as
\begin{equation}
P = - \frac{\sigma_{xx} + \sigma_{yy} + \sigma_{zz}}{3}
\end{equation}

The Mohr-Coulomb failure criterion outlined above may be normalised with respect to the cohesive strength. Similarly to the stress-based crack driving force criterion in \cref{eq:Dd}, this gives a crack driving force for pressure-dependent fractures:
\begin{equation} \label{eqn:Crack_Driving}
D_\textrm{d} = \left<\frac{\sqrt{\mu^2 + 1} \; \tau_{\textrm{max}} - \mu P }{\tau_{\textrm{c}}} \right>^2
\end{equation}
We note the absence of the $-1$ term acting as a damage threshold in \cref{eqn:Crack_Driving} when comparing the crack driving force found in \citet{Miehe2015a}, and \cref{eq:Dd}. As a result, the obtained phase field solutions will have a smooth transition between the damaged and non-damaged areas, in contrast to the more abrupt transition present for phase-field models with the threshold. Without the threshold, the model will predict failure with a more progressive softening, spreading the failure over several time increments and thereby aiding the numerical convergence. However, based on prior experience, the final failure outcomes will be roughly the same with and without the threshold.

\begin{figure}
    \centering
    \includegraphics[width = 0.5\textwidth]{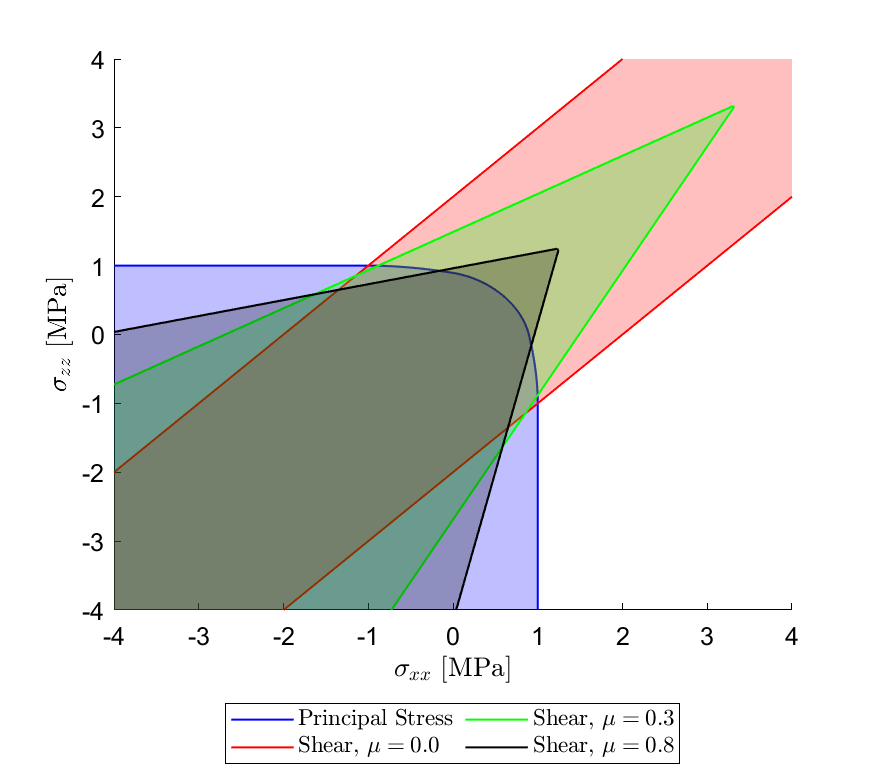}
    \caption{Diagram showing yield surfaces for the principal stress criterion (blue surface) and Mohr-Coulomb failure criterion for internal friction $\mu = 0.0 \;$ (red surface), $\mu = 0.3 \;$ (green surface), $\mu = 0.8\;$ (black surface). Shaded regions indicate combinations of stress $\sigma_{xx}$ and $\sigma_{zz}$ where the material will not undergo yielding (i.e. $D_\textrm{d} = 0$).}
    \label{fig:Yield_Surface}
\end{figure}

The yield surfaces for the Mohr-Coulomb based crack driving forces described in \cref{eqn:Crack_Driving} are presented in \cref{fig:Yield_Surface} considering $\mu = [0.0, \; 0.3, \; 0.8]$. For the no friction case ($\mu = 0.0$), a Tresca-type yield surface is produced such that the failure stress is solely due to the maximum shear stress $\tau_\textrm{max}$. In the principal stress space ($\sigma_1, \sigma_2, \sigma_3$), this gives a hexagonal prism of infinite length centered around the line $\sigma_{1} = \sigma_{2} = \sigma_{3}$ and the material is stable in regions where normal stresses are approximately equivalent (i.e. $\tau_\textrm{max}$ in \cref{eq:tau_max} tends to zero for $\sigma_{xx} \approx \sigma_{zz}$ and $\sigma_{zz} \approx 0$). The yield surface expressed in \cref{fig:Yield_Surface} shows two parallel lines of infinite length, representative of the longitudinal section of the 3D hexagonal prism where failure is independent of isotropic pressure. As a result, cracks can only nucleate when deviatoric (shear) stresses are present, whereas when ice undergoes hydrostatic (equi-triaxial) tension, it does not fracture. These deviatoric stresses also determine the plastic deformations through Glen's law (\cref{eq:GlenLaw}). Depending on the strain rate, the bulk strain energy density 
is dissipated by the deviatoric stresses at slow strain rates and through fracture at fast strain rates. 

As the value of internal friction increases, failure becomes dependent on isotropic pressure, tending towards a Mohr-Coulomb failure surface. In the 3D space of principal stresses, this is represented by a hexagonal-based pyramid, with the apex located on the $\sigma_{1} = \sigma_{2} = \sigma_{3}$ line and the critical applied stress $\sigma_\textrm{a}$ being equal to $\tau_\mathrm{c}/\mu$, thus for $\mu = 0$, $\sigma_\mathrm{a} = \infty$. This surface allows for ice to fail under both tensile and shear stress states, with ice being less likely to crack if hydrostatic tension is applied compared to the principal stress-based criterion, and allowing fracture before the tensile strength is reached when under compression. As a result, this model will allow fracture to occur in compressive regions based on deviatoric stresses, something which is not captured by the principal stress-based phase field formulation. In the following sections, we will test this new Mohr-Coulomb-based formulation considering idealized rectangular glaciers. 

\subsection{Basal Boundary Conditions}
For the interactions between the glacier and the bedrock, we consider a variety of boundary conditions represented in \cref{fig:Boundary_Conditions_MICI}, to test for cliff failure. In every case, we consider displacement in the vertical direction to be restrained, preventing the glacier from interpenetrating the basal rock. However, we vary the degree of motion in the horizontal direction; for most cases, we consider a Weertman-type sliding law which applies a basal shear traction $\tau_{\mathrm{b}}$ to oppose motion \citep{Bassis2021}:
\begin{equation} \label{eqn:Weertman_Sliding}
\tau_{\mathrm{b}} = -\left[ \frac{1}{C \left| \dot{\mathbf{u}}_{\text{t}}\right|^{1/m-1}} + \frac{\left|\dot{\mathbf{u}}_{\text{t}}\right|}{\tau_{0}}\right]^{-1} \dot{\mathbf{u}}_{\text{t}}
\end{equation}
This friction is dependent on the basal friction coefficient $C$, the friction exponent $m$ and the tangential sliding velocity $\dot{\mathbf{u}_{\text{t}}}$. Values of basal friction coefficient vary throughout Antarctica and are inferred through inversions of observed velocities \citep{MacAyeal1993, Barnes2022}; therefore, we consider a range of values for basal friction coefficient $C=10^5-10^9\;\mathrm{Pa}\;\text{m}^{-1/n}\text{s}^{1/n}$. The extreme cases of no friction and a fully frozen boundary are also considered: the free-slip basal boundary condition (\cref{fig:Boundary_Conditions_MICI}a) freely allows for horizontal displacement at the base and the no-slip condition (\cref{fig:Boundary_Conditions_MICI}c) does not allow horizontal displacement at the base ($u_{x} = 0$), representing a glacier with a frozen base.

\begin{table}
    \centering
    \begin{tabular}{c|c c}
        Material Parameter & Value & Units  \\
        \hline
         Young's Modulus $E$ & 9500 & MPa \\
         Poisson's Ratio $\nu$ & 0.35 & - \\
         Density of Glacial ice $\rho_\textrm{i}$ & 917 & $\textrm{kg m}^{-3}$ \\
         Density of Ocean water $\rho_\textrm{s}$ & 1020 & $\textrm{kg m}^{-3}$ \\
         Creep exponent $n$ & 3 & -  \\
         Creep coefficient $A$  & 7.156 $\times 10^{-7}$ & $\text{MPa}^{-\text{n}}\text{s}^{-1}$ \\
        Internal Friction $\mu$ & 0.8 & - \\
        Cohesive Strength $\tau_{\textrm{c}}$ & 1 & MPa \\
        Friction exponent $m$ & 3 & -  \\
        Reference Traction $\tau_0$ & 0.75 & \text{MPa} \\
        Phase Field Viscosity $\eta$ & 33.8 &  $\text{s}$ \\
        Phase Field Length Scale $\ell$ & 10 & \text{m}
    \end{tabular}
    \caption{Characteristic material properties for glacial ice assumed in this work (unless otherwise stated).}
    \label{tab:Properties}
\end{table}

\section{Land Terminating Glaciers}

We consider an idealised rectangular grounded glacier sitting on bedrock with thickness $H$ and length $L$, as shown in \cref{fig:Boundary_Conditions_MICI}. By adopting a length-to-thickness ratio of $L/H=6$, we ensure that for thicker glaciers the glacier length is long enough to capture both the near-terminus and far-field stress states. For the cases considered here, we assume a land-terminating glacier, $h_\text{w}=0$. We employ the plane strain assumption, as the out-of-plane dimension is typically larger than the in-plane length, allowing the three-dimensional geometry to be reduced to a two-dimensional geometry of flow-line with x and z denoting the horizontal (along flow) and vertical coordinates, respectively. The upper surface representing the air-ice interface is considered as a free surface and the displacement normal to the far left terminus is restrained to prevent rigid body motion in the horizontal direction. The material properties for glacial ice used within this study are reported in \cref{tab:Properties} unless stated otherwise. 

\subsection{Stress Distributions} \label{sub:Stress_Distribution}

\begin{figure*}
    \centering
   \includegraphics[width = \textwidth]{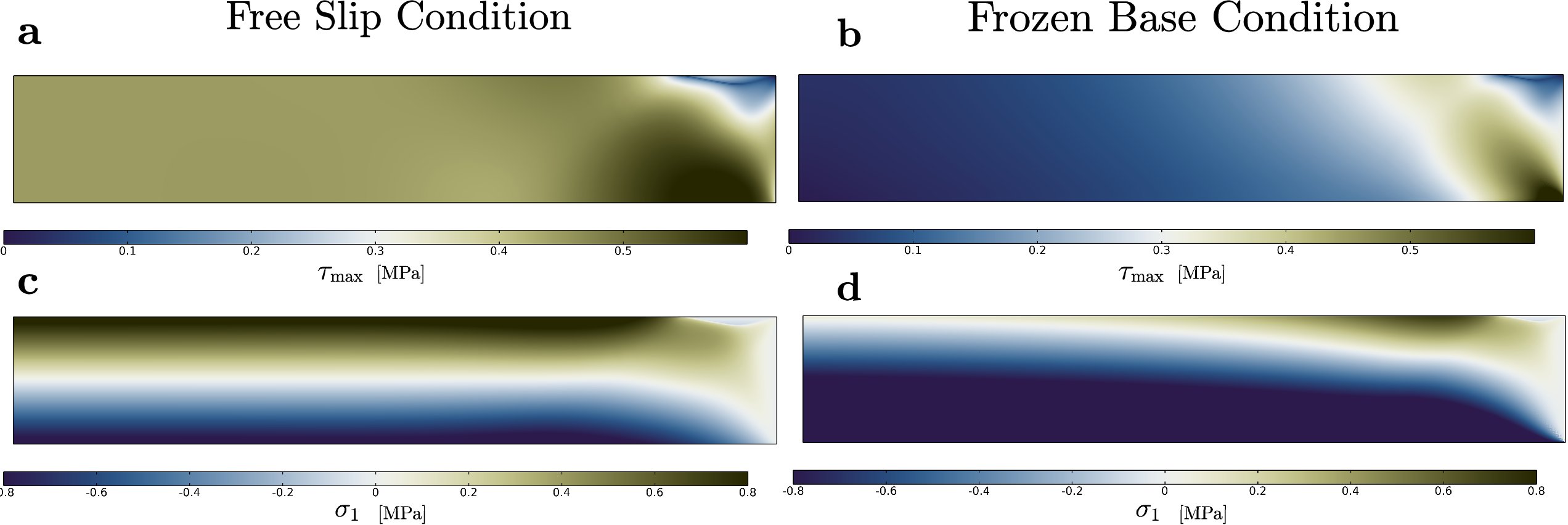}
   \caption{Steady state creep stress states showing maximum shear stress and first principal stress for a grounded glacier of height $H = 200 \; \text{m}$ undergoing free slip (a) and (c) and no slip (b) and (d)}
\label{fig:Stress_States_2}
\end{figure*}

Prior to conducting phase field damage simulations, the stress states in the pristine grounded glacier are considered. These stress states are obtained through time-dependent creep simulations without any damage, obtaining a steady state stress profile in the domain after 7 simulation days. We plot contour maps for the maximum shear stress $\tau_{\mathrm{max}}$ and maximum principal stress $\sigma_{\mathrm{1}}$ in \cref{fig:Stress_States_2} calculated using \cref{eq:tau_max,eq:sigma_1} respectively. For this illustration, we consider the extreme cases of frictional sliding at the base, namely free slip and frozen base for a land-terminating glacier of thickness $H = 200$ m.

For the free slip condition, the maximum shear stress $\tau_{\mathrm{max}}$ is plotted in \cref{fig:Stress_States_2}a, which is nonzero throughout the entire domain. In the far-field region, it is invariant with depth, having an approximate value of $\frac{1}{4}\rho_{\mathrm{i}} g H$. The maximum shear stress is greatest at the base of the glacier close to the front, with a peak value of 1.4 times the maximum shear stress in the far field region. The stress distribution for the maximum principal stress in the free slip glacier is plotted in \cref{fig:Stress_States_2}c. The upper surface layers in the far field region are subject to tensile stress, with a maximum value of $\frac{1}{2}\rho_{\mathrm{i}} g H$ being observed in the absence of ocean-water pressure. Maximum principal stresses vary linearly with depth and become compressive at the base, with the distribution being symmetrical about the centre line $z = \frac{H}{2}$. An edge effect is observed close to the front due to the traction-free condition, which leads to non-uniform (spatially varying) stress fields. Based on these stress profiles, it is expected that densely spaced crevasses will develop in the far-field region, while near the terminus, the upper surface is unlikely to develop any crevasses. If cracks were to propagate based on principal stresses, no cracks would develop at the base due to the compressive stress state, even near the terminus. In contrast, using a shear-based criterion allows for basal cracks to develop near the terminus due to the increase in shear stress near the base.  

The maximum shear stress for the frozen base glacier is presented in  \cref{fig:Stress_States_2}b. Values of maximum shear stress away from the glacier front are negligible, however a concentration in shear stress is observed at the base of the glacier near the terminus, with a maximum value of $1.35 \; \textrm{MPa}$. The maximum principal stress for the frozen base is shown in \cref{fig:Stress_States_2}d. In contrast to the free slip condition, the stress state is predominantly compressive in the far field region, with linear variation with depth and upper surface layers exhibiting smaller tensile stress (approximately $0.1 \; \textrm{MPa}$) compared to the free-slip case. Maximum values of principal stress are observed at a distance of one thickness $H$ from the glacier terminus at the upper surface. As a result of this stress state, surface crevasses are unlikely to initiate or propagate in the far-field region, whereas mode-I (tensile) cracks are likely to develop near the terminus at the surface, and mode-II (shear) cracks at the base. It is also observed that the maximum principal and shear stresses (see \cref{fig:Stress_States_2}) are proportional to the glacier thickness $H$, due to load contributions being gravitational body forces. This is the case regardless of basal boundary conditions.

\subsection{Cliff Failure - Influence of Basal Boundary Condition} \label{sec:BasalBC}

We next conduct time-dependent phase field damage simulation studies to explore the conditions enabling ice cliff failure. The steady state stress states in creeping glaciers reported in \cref{fig:Stress_States_2} are used to initialise the phase field simulations, so that the propagation of damage is governed by the incompressible viscous rheology rather than the compressible linear elasticity. It should be noted that while the viscous stresses are used for the initial stress distribution, as the fractures propagate the stress state within the computational model will instantaneously adapt to the presence of these new fractures through linear-elastic deformations, with these updated stresses then driving further fracture propagation and causing further viscous relaxation over longer times.

\begin{figure*}
    \centering
    \includegraphics[width = \textwidth]{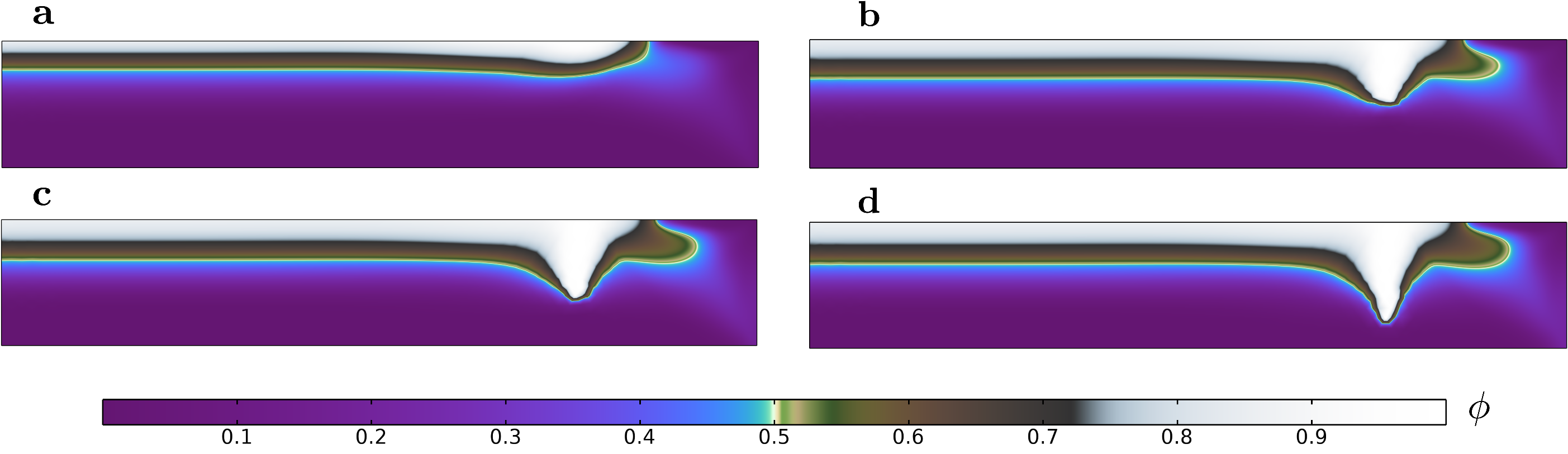}
    \caption{Phase field damage evolution over time for a land terminating glacier of height $H = 200$ m undergoing free slip at the base with internal friction $\mu = 0.8$ and cohesion $\tau_{\textrm{c}} = 1 \; \textrm{MPa}$ at time (a) t = 15 s, (b) t = 50 s, (c) t = 75 s and (d) t = 200 s}
    \label{fig:Shear_Failure_PF_Free_Slip_H200_0.8}
\end{figure*}

Phase field contour plots for the grounded glacier undergoing free slip are presented in  \cref{fig:Shear_Failure_PF_Free_Slip_H200_0.8}, with $\phi=0$ (blue) indicating the ice is undamaged and $\phi=1$ (white) indicating it is fully fractured. Uniform damage initiates on the upper surface in the far field region and stabilises at a thickness of approximately $0.5H$, a depth which is consistent with the Nye zero stress prediction for a land-terminating glacier. A concentration of damage is located close to the ice front and propagates vertically downward to a normalised depth of $0.76H$. This difference in crevasse depth is a result of boundary effects and crack shielding, with the right-most crevasse propagating to a depth comparable to that predicted for a single crevasse in isolation, whereas the remainder of the crevasses behave as densely spaced, thus following the zero-stress depth estimates. The damage accumulated in the free slip glacier is driven by the longitudinal stress and as a result can be categorised as mode I tensile failure. It is acknowledged that the damage presented in \cref{fig:Shear_Failure_PF_Free_Slip_H200_0.8} is not localised to produce sharp densely spaced crevasses, instead producing a uniform damage region in the upper surface to represent a field of closely spaced crevasses away from the terminus. It is possible to overcome this, for instance by imposing a crack driving force threshold and inserting rectangular notches to localise damage to propagate directly beneath pre-existing cracks, as sharp mode I fractures \citep[see][]{Clayton2022}. However, this requires inserting notches beforehand to cause the surface crevasses to properly localise, thus removing the ability to study where and if these crevasses nucleate. As such, the method used here will produce smeared damage regions to indicate the presence of crevasse fields, while it does capture their nucleation starting from a pristine ice sheet.  

\begin{figure*}
    \centering
    \includegraphics[width = \textwidth]{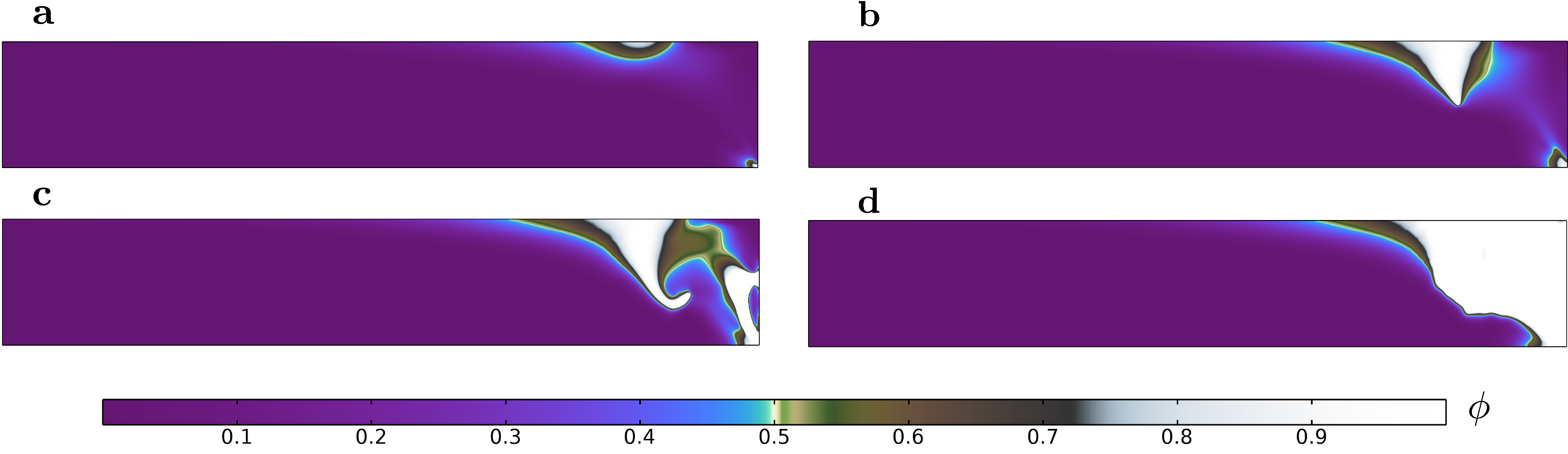}
    \caption{Phase field damage evolution over time for a land terminating glacier of height $H = 200$ m subject to a frozen base with internal friction $\mu = 0.8$ and cohesion $\tau_{\textrm{c}} = 1 \; \textrm{MPa}$ at time (a) t = 15 s, (b) t = 60 s, (c) t = 150 s and (d) t = 250 s}
\label{fig:Shear_Failure_PF_Frozen_H200_0.8}
\end{figure*}

The phase field contour plots for the frozen base glacier are reported in  \cref{fig:Shear_Failure_PF_Frozen_H200_0.8}. In contrast to the free-slip results, the damage is localised at the base of the glacier near the terminus and in the upper surface regions approximately one thickness away from the terminus. Damage at the upper surface initially propagates downwards, and at greater depths, the fracture path begins to curve towards the glacier front. Simultaneously, damage initiates at the base near the terminus as a result of the concentration in maximum shear stress $\tau_{\mathrm{max}}$ reported in  \cref{fig:Stress_States_2}b. This basal fracture propagates upwards in a mixed-mode manner and cliff failure is observed once the basal fracture approaches the surface fracture and makes contact with the terminus. Next, in \cref{fig:Shear_Failure_PF_Frozen_H200_0.8}c we can see these two crevasses propagating in the glacier, and a third crack appears: the first first from the surface propagating downward due to the tensile stress concentration near the surface; the second from the right bottom corner propagating upward due to the stress concentration from the fixed boundary condition, and the newly formed third crack from the right edge propagating horizontally due to the stress changes induced by the other two cracks. These then coalesce, which causes wide spread damage and mass loss observed in  \cref{fig:Shear_Failure_PF_Frozen_H200_0.8}d. 
Fracture coalescence occurs in a rapid brittle manner, and once cliff failure is achieved, a stable cliff surface is observed. After these initial fractures, no further glacier retreat occurs because the bed is flat and the fracture face has a low-grade slope, thus the exposed surface is much more slanted than the initial upright cliff surface. 

\begin{figure}
    \centering
    \includegraphics[width = 0.5\textwidth]{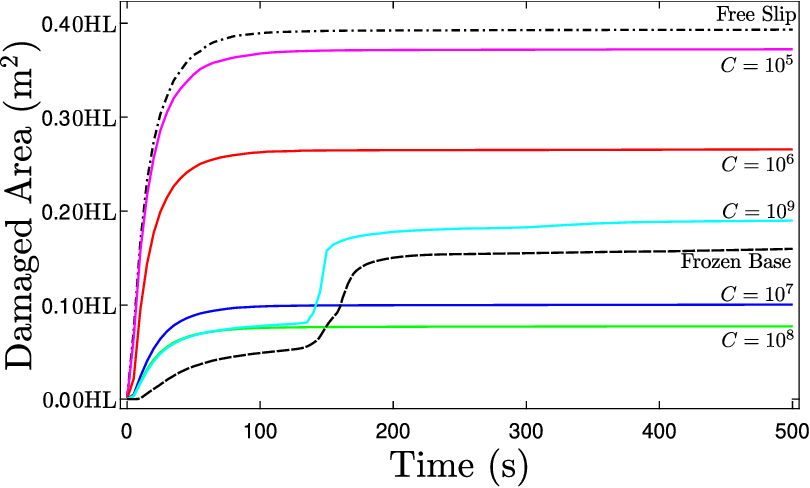}
    \caption{Graph showing damage accumulation area normalised with respect to in-plane glacier area ($H \times L$) versus time for a land terminating graph of height $H = 200 \; \textrm{m}$ for different basal boundary conditions.}
    \label{fig:Dam_Accumulation}
\end{figure}

The damage accumulation, $\int_\Omega \phi\;\text{d}\Omega$, versus time is shown in \cref{fig:Dam_Accumulation}, normalised with respect to the in-plane glacier area ($H \times L$). Note that the damage accumulation defined above does not indicate the area of ice lost but rather indicates the area of crevassed/damaged regions; for example, the presence of the crevasses near the surface does not result in iceberg calving as evident from  \cref{fig:Shear_Failure_PF_Free_Slip_H200_0.8}. However, the jumps in \cref{fig:Dam_Accumulation} indicate that areas of the ice sheet become detached, such as that observed for the frozen base case between $150-250\;\text{s}$ in \cref{fig:Shear_Failure_PF_Frozen_H200_0.8}c-d. Here, we consider the frozen and free-slip cases, as well as the glacier subject to basal shear, with a variety of basal friction coefficient values $C$. For basal friction coefficients $C<1\times 10^5$, the basal shear stress is sufficiently small that damage accumulation tends towards the free slip glacier case. Damage accumulation occurs in a rapid brittle manner, with damage accumulation areas stabilising at approximately $t = 50$ s. As the basal friction coefficient increases, basal shear resists glacial flow and the total damage accumulation decreases due to the longitudinal stress profile becoming more compressive, in turn reducing the depth of surface crevasses.  For high values of $C$, far-field damage is no longer present, and damage only accumulates close to the glacier front. Cliff failure is observed for basal friction coefficients greater than $C>1\times 10^9$, with this behaviour tending towards the frozen base case. 

\subsection{Cliff Failure - Influence of Internal Friction}

We explore the influence of internal friction parameter $\mu$ on the mode of fracture, as there is a wide range of reported values in the literature. 
\citet{Weiss2009} conducted biaxial compression tests on columnar ice and determined friction coefficient to be scale independent with an approximate value of $\mu = 0.8$. By contrast, \citet{Bassis2012} conducted cliff failure analysis by considering $\mu = 0$, reverting to a Tresca yield criterion. To consider the effect of the choice of friction parameter, phase field fracture simulations are performed for the frozen base land terminating glacier case with height $H = 200 \; \text{m} $ and consider the extreme values of $\mu = 0$ and $\mu = 0.8$ that were reported in the literature. Results for $\mu = 0.8$ have been presented in  \cref{fig:Shear_Failure_PF_Frozen_H200_0.8} and discussed previously.

\begin{figure*} 
    \centering
    \includegraphics[width = \textwidth]{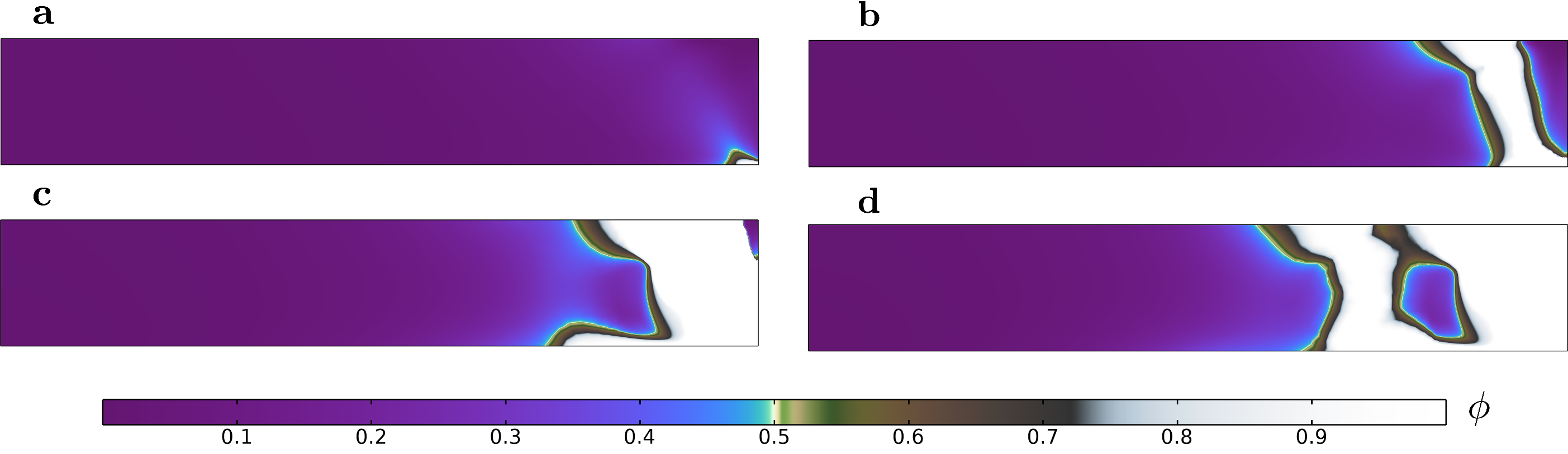}
    \caption{Phase field damage evolution over time for a land terminating glacier of height $H = 200$ m subjected to a frozen base with internal friction $\mu = 0.0$ and cohesion $\tau_{\textrm{c}} = 1 \; \textrm{MPa}$ at time (a) t = 17 s, (b) t = 35 s, (c) t = 70 s and (d) t = 93 s}
    \label{fig:Shear_Failure_PF_H200_0.0}
\end{figure*}

We present the phase field contour plots for the no-friction case in  \cref{fig:Shear_Failure_PF_H200_0.0}. The absence of the isotropic pressure $P$ in the crack driving force results in damage initiating solely as a result of the maximum shear stress $\tau_{\textrm{max}}$, leading to a Tresca-type yielding criterion. Maximum shear stress is present at the base (cf. \cref{fig:Shear_Failure_PF_H200_0.0}a) of the glacier in close proximity of the front, causing a fracture to propagate from the base upwards and penetrate the full thickness of the glacier (cf. \cref{fig:Shear_Failure_PF_H200_0.0}b). As ice detaches from the front of the ice sheet, the damage region widens and a new basal fracture propagates along the frozen base (cf. \cref{fig:Shear_Failure_PF_H200_0.0}c), detaching the ice from the bedrock. Once this basal crack is sufficiently long (with a length comparable to the ice thickness), the damage grows upwards and a second crack propagates through the entire glacier thickness \cref{fig:Shear_Failure_PF_H200_0.0}d). This suggests crack propagation to be a recurring event with the new glacier front created after each time a crevasse penetrates full ice thickness, until it disintegrates the entire glacier domain in this simulation. This ice cliff instability seems to be a result of the steep cliff face created after each time the front detaches and the stress state reconfigures. However, despite the differences in failure patterns between the high and low internal friction cases (i.e. $\mu=0$ and $\mu=0.8$) in \cref{fig:Shear_Failure_PF_Frozen_H200_0.8} and \cref{fig:Shear_Failure_PF_H200_0.0}, we find that a land terminating no-slip glacier of height $H = 200 \; \textrm{m}$ is prone to observe ice cliff failure. This means that basal sliding friction (depending on glacier geometry and subglacial hydrology) is a more significant factor in determining ice cliff failure than internal friction (depending on material behavior). 

\subsection{Cliff Failure - Influence of Cohesive Strength}

There is also a large variation in reported values of cohesion $\tau_{\mathrm{c}}$ in the literature. Firstly, \citet{Beeman1988} reports a cohesion value of $\tau_\mathrm{c} = 1$ MPa under low confining pressures for cold ice. This value of cohesion has been used in several numerical studies such as those by \citet{Bassis2012} and \citet{Schlemm2019}. However, observational data suggests lower values of cohesion closer to $\tau_\mathrm{c} = 0.5$ MPa might be more appropriate to capture realistic failure criteria \citep{Vaughan1993}. \citet{Frederking1988} reports similar values of cohesion, with an average value of $\tau_\mathrm{c} = 0.6$  $\textrm{MPa}$ being obtained from laboratory testing. By contrast, triaxial tests conducted by \cite{Rist1994, Gagnon1995} reported values of shear strength of up to $5$  MPa.

\begin{figure}
    \centering
    \includegraphics[width = 0.45\textwidth]{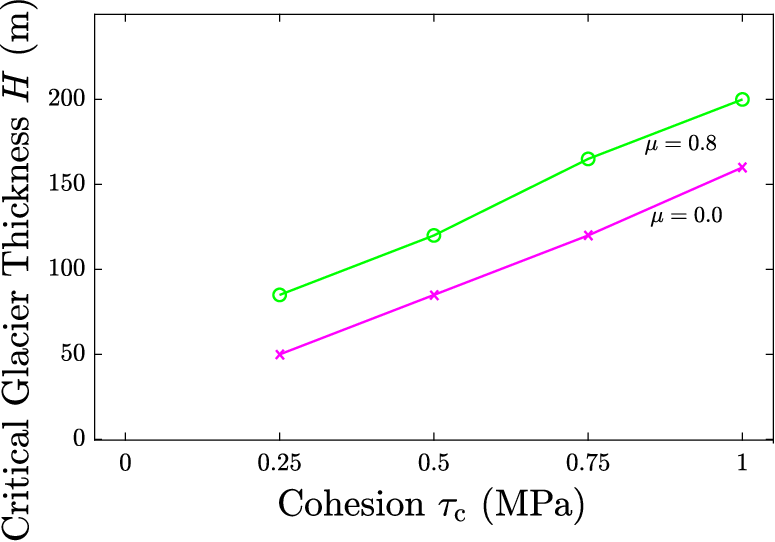}
    \caption{Minimum glacier thickness required to trigger cliff failure for several values of  $\tau_\mathrm{c}$ for $\mu = 0.0$ and $\mu = 0.8$ for a frozen-base, land terminating glacier.}\label{fig:Cliff_Height_vs_Strength}
\end{figure}

We, therefore, run phase field damage simulations for cohesion $\tau_\mathrm{c} = [0.25,\; 0.5,\; 0.75,\; 1]\;\text{MPa}$ for the frozen base case at different glacier thicknesses, to determine the minimum height at which cliff failure is observed, assuming internal friction $\mu = 0.8$. Based on the crack driving force $D_\textrm{d}$ in \cref{eqn:Crack_Driving}, an alteration in cohesion will scale with the magnitude of the crack driving force. This will not alter the observed failure pattern, although cohesion will influence the minimum or critical height at which cliff failure occurs, as shown in \cref{fig:Cliff_Height_vs_Strength}.

For $\tau_\mathrm{c} = 1$ MPa, $\mu = 0.80$, land terminating glaciers of height $H \geq 200$ m are subject to cliff failure, which is consistent with the $H = 220$ m estimation by \citet{Bassis2012}. As the cohesion decreases, the height at which cliff failure occurs reduces, lowering to $H \geq 85$ m when considering $\tau_\mathrm{c} = 0.25$ MPa, as shown in \cref{fig:Cliff_Height_vs_Strength}. These results are obtained by performing simulations for a range of ice thicknesses, determining the minimum thickness required for failure with an accuracy of $5\;\text{m}$. Notably, for the range of cohesion values considered, the relation between cohesion and stable cliff height is close to linear. However, it is expected that as the cohesion approaches zero, the stable cliff height will approach zero thickness. A similar relationship between cliff height and cohesion is observed for the zero internal friction case ($\mu = 0.0$), however, the critical thickness is reduced by $40 \; \textrm{m}$ compared to $\mu = 0.80$. As stable cliff-heights observed are typically in the range of $100\;\text{m}$ \citep{Scambos2011}, we can conclude that for land terminating glaciers realistic values for the cohesion are in the range of $\tau_c=0.25-0.5\;\text{MPa}$ for $\mu=0.8$ (cf. \cref{fig:Cliff_Height_vs_Strength}).

\section{Ocean Terminating Glaciers}

We next turn our attention to thicker glaciers that terminate at the ocean. A depth-dependent hydrostatic ocean-water pressure $p_\textrm{w}$ is applied at the far right terminus of the glacier with a magnitude of:
\begin{equation}
    p_\textrm{w} = \rho_\textrm{w} g \left< h_\mathrm{w} - z \right>
\end{equation}
to represent an ocean with height $h_\text{w}$. The inclusion of this ocean-water pressure provides a compressive stress which resists glacier motion and allows for thicker glaciers to stabilise. As the cliff failure was most pronounced for frozen base conditions, these will be used here. We note that this might not be the most realistic boundary condition, as ocean-terminating glaciers typically are not frozen to the base, instead having a wide range of friction coefficients. However, as will also be discussed later on, the presence of an ocean water pressure at the terminus restricts crevasse growth at the base, instead causing the failure to occur solely due to surface crevasses. As a result, the impact of the basal sliding conditions is lessened in these cases. Additionally, we do not consider any flotation/uplift throughout this work \citep{Trevers2019}, instead solely considering ice sheets with sufficient free-board to remain grounded.  

\begin{figure*} 
    \centering
    \includegraphics[width = \textwidth]{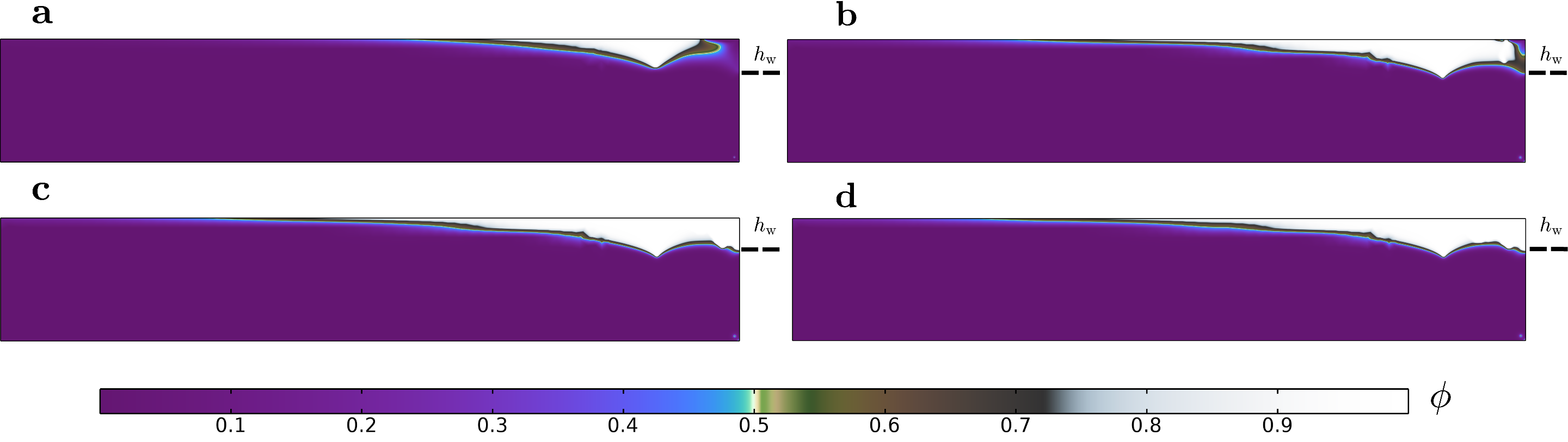}
    \caption{Phase field damage evolution over time for an ocean terminating grounded glacier of height $H = 800$ m and ocean-water height $h_\mathrm{w} = 585$ m subject to a frozen base with internal friction $\mu = 0.8$ and cohesion $\tau_{\textrm{c}} = 0.5 \; \textrm{MPa}$ at time (a) t = 5 s, (b) t = 40 s, (c) t = 100 s and (d) t = 250 s}
\label{fig:Phase_Field_Oceanwater}
\end{figure*}

The phase field contours for an ocean terminating glacier of height $H = 800 \; \textrm{m}$ and ocean-water height $h_\textrm{w} = 585 \; \textrm{m}$ are presented in \cref{fig:Phase_Field_Oceanwater} for a value of $\tau_\textrm{c} = 0.5 \; \textrm{MPa}$. Damage is localised in the upper surface close to the calving front and slumping is observed in the ice above the waterline until a subaerial calving event is achieved. By contrast to the land terminating case, full thickness failure is not achieved. Instead, a stable ice thickness at the calving front is sustained, equal to the ocean-water height $h_\textrm{w}$. The retreat of the glacier as a result of subsequent buoyant calving is not considered, as this requires considering the uplift due to the buoyancy force and melt-undercutting. As a result, no mechanisms are included for water to reach underneath the ice-sheet once the glacier foot is exposed, and no basal cracks can be created post-slumping as the calving front becomes buoyant. In the remainder of this section, we will refer to this type of calving as cliff-failure. 

Cliff failure in ocean-terminating glaciers is therefore dependent upon the exposed free-board above the ocean waterline (i.e. $H - h_\textrm{w}$), not on the thickness of the ice itself. This behaviour pattern is consistent with the results of \citet{Parizek2019} and also supports empirical calving laws, based on the height above buoyancy \citep{VanDerVeen1996}. For the presented results, the minimum glacier free-board to cause cliff failure is $215 \; \textrm{m}$, which is larger than the critical glacier thickness for the land terminating case ($H = 125 \; \textrm{m}$). This increase in critical free-board of marine-terminating glaciers is likely observed due to the absence of damage at the base, due to the stabilisation effect of the ocean-water pressure $p_\mathrm{w}$. 

\begin{figure*}
    \centering
    \includegraphics[width = 1.0\textwidth,clip=true,trim={50 0 50 0}]{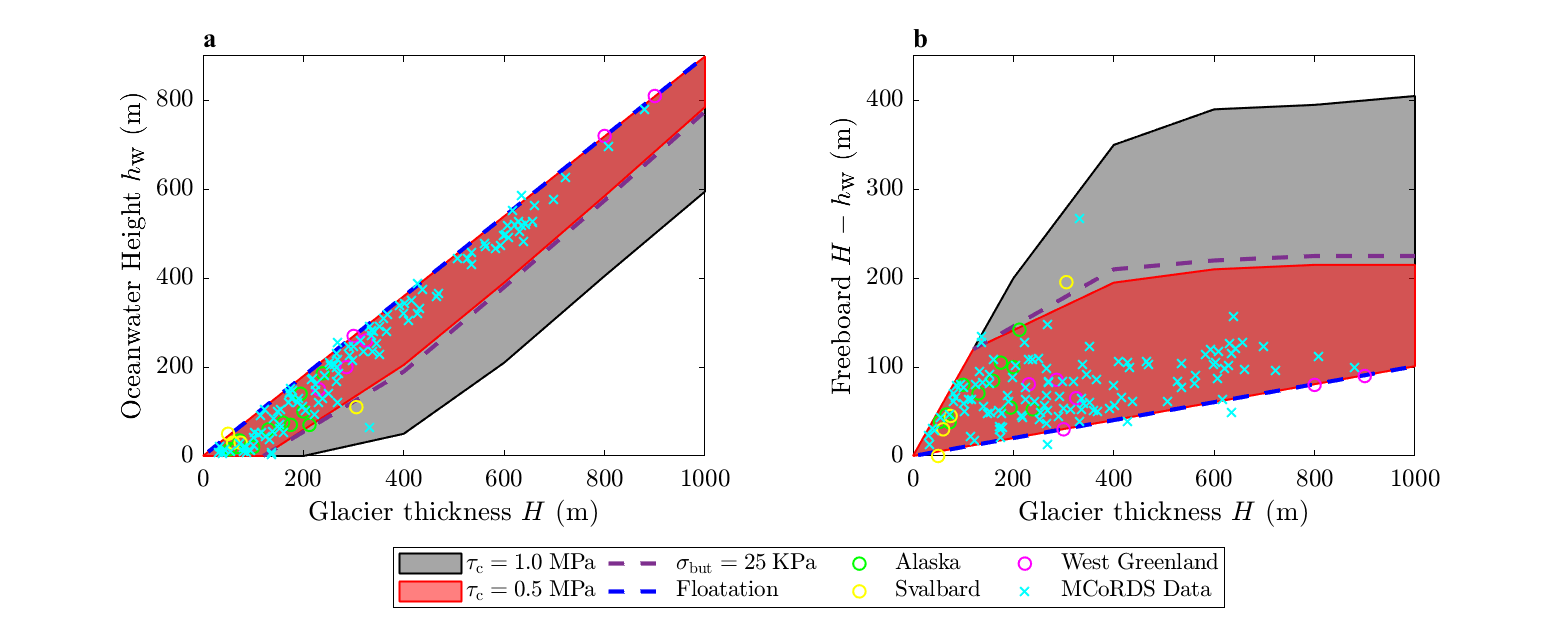}
\caption{Combination of glacier thickness and (a) ocean-water height or (b) freeboard required for stable ice cliffs to exist (shaded regions), flotation to occur (blue dashed line), or cliff slumping to trigger (exceeding the shaded area). Observational data from Alaska, Svalbard and West Greenland Glaciers from \citet{Pelto1991} and MCoRDS radar data for various Greenland outlet glaciers from \citet{Ma2017}.}
\label{fig:Cliff_Height_vs_Oceanwater}
\end{figure*}

Similar to the study into the effects of cohesion, we conduct another parametric study with multiple phase field fracture simulations for a variety of glacier thicknesses and ocean-water heights, producing the stability envelope shown in \cref{fig:Cliff_Height_vs_Oceanwater}a. This shows the critical value of ocean-water height $h_\mathrm{w}$ required to cause ice cliff freeboard failure at glacier thickness increments of 200 m. If the ocean water exceeds this critical value, (i.e. the data point lies within the shaded region) calving will not be observed, however if the ocean water is below the critical value (i.e. the data point lies outside the shaded region and below the stability envelope), freeboard cliff failure will occur. An upper bound for the stability envelope for grounded glaciers (blue dashed line in \cref{fig:Cliff_Height_vs_Oceanwater}a) is determined by the ocean-water height required for the glacier to become buoyant, (i.e. $h_\mathrm{w} = \left( \rho_\mathrm{i}/\rho_\mathrm{w} \right)H$).

A mostly linear relationship between glacier thickness and critical ocean-water height is observed for thicknesses exceeding $H=400$ for $\tau_c = 0.5$ MPa or $H=600$ m for $\tau_c = 1$ MPa, meaning the critical glacier free-board $(H-h_\mathrm{w})$ is independent of glacier thickness as shown in \cref{fig:Cliff_Height_vs_Oceanwater}b. Instead, the critical glacier free-board is dependent on cohesion $\tau_\mathrm{c}$. For $\tau_\mathrm{c} = 0.5\;\textrm{MPa}$, the stability region is bounded by the red zone in \cref{fig:Cliff_Height_vs_Oceanwater}. We observe a critical glacier free-board of $H-h_\mathrm{w} \approx 215 \; \text{m}$ which is in accordance with observations (i.e. the scatter data points all fall within the red stability zone). However, increasing the cohesion to $\tau_\mathrm{c} = 1.0$ MPa gives a larger stability region (grey zone in \cref{fig:Cliff_Height_vs_Oceanwater}) and results in a critical glacial free-board of $H-h_\mathrm{w} \approx 405 \; \text{m}$. Observational measurements for glacier thickness and ocean-water height have been added to \cref{fig:Cliff_Height_vs_Oceanwater}. The data used was recorded by Landsat 4, reported in \citet{Pelto1991}, with measurements taken from Columbia, Alaska, West Greenland and Svalbard glaciers. Additional data for Greenland glaciers including the Helheim, Jakobshavn, Petermann and Hayes glaciers has been measured using  Multichannel Coherent Radar Depth Sounder (MCoRDS) from \citet{Ma2017} (cyan datapoints in \cref{fig:Cliff_Height_vs_Oceanwater}). When comparing the observational data to the stability envelope produced from phase field simulations, it can be seen that the majority of observations are encompassed within the stability envelope for $\tau = 0.5 \; \textrm{MPa}$, whereas  $\tau = 1.0 \; \textrm{MPa}$ provides an overly conservative approximation. The majority of glacier observations in \cref{fig:Cliff_Height_vs_Oceanwater} show thicknesses of less than $400 ~\textrm{m}$, whilst thicker glaciers tend towards buoyancy. 

The blue dashed line in \cref{fig:Cliff_Height_vs_Oceanwater} shows the existence of a lower limit for grounded glacier ice thickness $H$, which corresponds to the cases where the critical ocean-water height for cliff failure exceeds the ocean-water height for the glacier to become buoyant ($ h_\textrm{w}^\textrm{crit} \geq \frac{\rho_\mathrm{i}}{\rho_\mathrm{sw}}H$). This means that glaciers thicker than this limit will have to form a floating ice shelf. For floating ice tongues and ice shelves, failure by shear is unlikely to occur due to the no-slip condition being replaced with a buoyancy pressure at the base. Instead, failure is dictated by the propagation of rifts leading to the detachment of tabular icebergs instead of cliff slumping \citep{Bassis2012,Huth2023}, a fracture mechanism which is not considered within this study This fracture mechanism is not considered within this study but can be captured by the proposed fracture model if the pore-water pressure were to be included.

\subsection{Inclusion of Buttressing Stresses}

The above analysis of cliff failure in marine-terminating glaciers neglected any buttressing stresses that may be applied at the glacier front due to lateral friction, resistance at pinning points or ice mélange formed by sea ice and previously detached icebergs. Prior simulations have shown that the presence of ice mélange leads to reductions in iceberg calving \citep{Krug2015,Robel2017}, including ice melange with ice shelf rifts \citep{Huth2023}. This may contribute to seasonal variations in calving rates (as well as increased production in meltwater), where mélange is present in winter periods and absent in the summer. To investigate the influence of buttressing stress on calving, a horizontal traction $\sigma_\textrm{but}$ is applied at the far right terminus of the tidewater glacier, with a magnitude of $25 \; \textrm{kPa}$ and a contact area $w$ extended $25 \; \textrm{m}$ above the waterline $h_\mathrm{w}$ and  $55 \; \textrm{m}$ beneath the waterline as per \citet{Bassis2021}.
 
The modified stability envelope for $\tau_\mathrm{c} = 0.5 \; \textrm{MPa}$ is represented by the purple dashed line in \cref{fig:Cliff_Height_vs_Oceanwater}, showing the negligible increase in stable heights obtained through buttressing stress for grounded glaciers. For this configuration of buttressing stress, we observe an increase of $10 \; \textrm{m}$ in the critical glacier free-board for glaciers thicker than $H \geq 600 \; \textrm{m}$, whilst for thinner glaciers, a $15 \; \textrm{m}$ increase in critical glacier free-board is observed. This result is expected, as the difference in ocean-water pressure can be approximately equated to the buttressing stress and thus the change in ocean-water height is $\Delta h_\mathrm{w} \approx \sqrt{\frac{2 \sigma_\textrm{but} w }{\rho_\mathrm{w} g}}$. Whilst this increase in free-board is negligibly small, further work is needed to better represent the effect of buttressing stress in our modeling study. 

\section{Discussion}
\subsection{Limitations of this modeling study}

We acknowledge that there are a few limitations to the assessment presented in this paper. The ice sheet geometry used is a highly idealised rectangular slab with a horizontal grounding line and no pre-existing crevasses. Originating from extensional strains or previous fracture events, pre-existing crevasses could act to localise damage by increasing stresses beyond what is expected for a pristine ice sheet. As a result, crevasses are more likely to propagate and the ice sheet might calve at greater ocean heights. The influence of a prograde or retrograde bed slope is also not explored within this paper, which may lead to progressive failure and rapid grounding retreat. While these progressive failures were observed for some of the $\mu=0$ cases, prograde slopes would reduce the stability of the ice cliff, whereas retrograde slopes would expose taller ice cliffs as subsequent crevassing occurs. While not considered here, the current model could be applied to study these events, as no a priori information is required for the expected location of crevasses, and intersecting crevasses are automatically resolved. 

We find that in order for cliff failure to occur in land-terminating glaciers, they must be undergoing high basal friction or are subject to no-slip frozen base conditions. This assumption is valid if basal ice is sufficiently cold, and friction coefficients approaching this no-slip condition exist \citep{Pollard2012}. However, basal lubrication, either through surface meltwater draining to the bed or melt at the bedrock due to frictional and geothermal heat may lead to increased sliding. This increase in sliding may favour the propagation of crevasses in far-field regions as opposed to cliff failure near the glacier terminus. 

In addition, the model also neglects the effects of melt undercutting and buoyancy pressures applied to the base of the glacier after a subaerial calving event, which may trigger subsequent buoyant calving and lead to rapid glacier retreat. Glacial mass losses in the form of basal undercutting may lead to altered stress states which would affect calving rates. However, it is unclear whether submarine melting prevents or enhances calving \citep{Ma2019,O'Leary2013}. We also have considered the ice as a solid (non-porous) material, such that the effective stresses driving cracks are based on the stresses in the ice itself. As a result of this assumption, the ice below the oceanwater line is fully loaded in compression, preventing any basal crevasses from forming. Potential pre-existing (water-filled) crevasses at the base are also neglected, limiting the propagation of basal hydro-fractures. A different assumption that could have been made would follow \citet{Ma2017}, where the ice is assumed to have sufficient basal crevasses such that the oceanwater pressure acts throughout the whole domain as if it is a porous material, instead of just applying a compressive pressure at the terminus. This would drastically reduce the compressive stresses (now carried by the oceanwater pressure instead of the solid stresses), and could allow for basal crevasses to develop.

The above analysis has assumed that all fractures are dry, so we did not consider the opening effect of water pressure in damaged regions. While the inclusion of meltwater is possible in the presented framework \citep[see e.g.][]{Duddu2020,Clayton2022}, we have not performed studies on its effect here to limit the number of parameters being varied. However, if included, the inclusion of meltwater-driven hydrofracture would result in increased tensile stresses in fractured regions and reduce the critical glacier free-board value to cause calving. Furthermore, as meltwater-driven cracks typically are mode-I (tensile) fractures, they tend to propagate vertically, causing steeper cliffs to be created compared to those predicted with our models. As such, including meltwater within this framework would allow for further studying the successive crevassing observed within our $\mu=0$ cases, \cref{fig:Shear_Failure_PF_H200_0.0}, providing insights into the role of meltwater in ice-cliff stability. Our stability envelope in Figure 10 wider compared to that of Bassis and Walker (2012), likely because we do not consider hydrofracture of basal crevasses. As a result our stability envelope only slightly converges towards floatation at higher ice thicknesses.

\subsection{Implications for fracture modelling}

The results produced in this research inform ice fracture modellers on appropriate values of cohesion and internal friction coefficients to use in future studies. As stated previously, a large amount of variation exists in the literature for experimental data for $\tau_\textrm{c}$ and $\mu$. It is found that for high values of internal friction $\mu=0.8$ and values of cohesion in the range of $\tau_\textrm{c}=0.3-0.6\;\text{MPa}$ provide a stability envelope for ice cliff failure which is in good agreement with observational measurements from \citet{Pelto1991} and \citet{Ma2017}. These values of cohesion are also in accordance with shear tests conducted by \citet{Butkovich1956,Paige1967,Frederking1988}. 

As variations in internal friction $\mu$ alter the fracture criterion significantly, this research has also indicated the role of extending phase field models to include the correct failure criteria for ice. This research has provided a first step in extending these methods beyond only considering tensile-driven mode I crevasses. However, further analysis is required into what the correct failure surface for mixed modes fractures within ice is. The produced failure events match to observed cliff heights, it is yet unclear if the Mohr-Coulomb type fracture criterion used here is most appropriate. Alternatives used within the literature for other materials include Drucker-Prager-type models \citep{Lorenzis2021,Navidtehrani2022}, or models which assign unique fracture energies for mode I and II \citep{Feng2023}. However, even though the exact fracture criterion to use is unclear, the results obtained here indicate that standard phase field schemes that are commonly used to model crevasses through ice sheets \citep[e.g. ][]{Sun2021,Clayton2022,Sondershaus2022}, are not well suited to capture slumping-type failure near the terminus region due to the fact they solely capture tensile failure. Instead, the Mohr-Coulomb yield surface-based crack driving force function presented here is capable of capturing mixed-mode fracture processes.

\subsection{Implications for cliff stability criteria}

Our results show that, for sufficiently thick glaciers, failure towards the waterline is independent of the total ice thickness but rather depends on the freeboard. While we do not model it here, if instead we assumed that once a crevasse reaches below the waterline and connects to the ocean itself, it could allow floatation to result in basal crevasses, and hydro-fractures will cause crevasses to connect between the base and the already-fractured ice. This assumption, made in waterline crevasse depth models \citep{Benn2007,Nick2010}, would extend our stability envelope from \cref{fig:Cliff_Height_vs_Oceanwater} from indicating when cliff slumping towards the waterline occurs to instead indicate when full-thickness failure occurs.

The stability diagram obtained from the phase field method (\cref{fig:Cliff_Height_vs_Strength}) shows a wider stability zone compared to that obtained from discrete element \citep{Benn2018} and analytical approaches \citep{Bassis2012}, which can be attributed to the inclusion of viscosity, cohesion and internal friction in our analysis. As a result, our phase field model predicts a critical free-board of approximately $200\;\text{m}$ for $\tau_\text{c}=0.5\;\text{MPa}$; whereas, the discrete element approach of \citet{Benn2018} predicts a maximum cliff height of $110 \; \textrm{m}$ for a strength of $\tau_\mathrm{c} = 1 \; \textrm{MPa}$. Thus, our prediction is within the range of cliff heights predicted by other works, for example,  \citet{Clerc2019} give a range of 90 m - 540 m depending on whether ice behaves as a purely elastic or viscous material; \citet{Crawford2021,Bassis2021} predict structural cliff failure of cliff heights greater than 135 m and \citet{Parizek2019} predict a range of 100 m - 245 m for threshold cliff height depending on the mechanical competency of ice. Notable, the criterion used in \citet{Pollard2015,DeConto2016} determines a limit for subaerial cliff heights of $100 \; \textrm{m}$ for a strength of $1 \; \textrm{MPa}$, by assuming that there are pre-existing surface and basal crevasses; whilst \citet{Bassis2012} predicts a cliff thickness of $221 \; \textrm{m}$ for land terminating glaciers with $\tau_\mathrm{c} = 1 \; \textrm{MPa}$ when considering no pre-existing crevasses.

Another finding of our study is that the threshold cliff height for failure for a land-terminating glacier is less than that for the freeboard cliff height of a marine-terminating glacier. Figure \cref{fig:Cliff_Height_vs_Oceanwater} suggests that failure is observed above a height of $125\;\text{m}$ in land terminating versus the height above the waterline of $200\;\text{m}$ for ocean-terminating cliffs. This is a result of the lack of basal crevasse propagation at the glacier toe in a marine terminating glacier, due to the compressive oceanwater pressure. 
As ice sheets retreat and glaciers lose their ice shelves and become land-terminating, this finding suggests lower thresholds for ice cliff failure, which should be accounted for in ice sheet models. Alternatively, if ice sheet retreat coincides with changes in basal friction coefficient this would impact threshold cliff heights as well. For example, if the ice-bed interface is well lubricated due to meltwater or ocean water at the base, then steep ice-cliffs of above $200\;\text{m}$ could be stable (\cref{fig:Shear_Failure_PF_Free_Slip_H200_0.8}), as the failure is dictated by tensile crevassing in the far field. Thus, accounting for the different failure modes due to shear and tension in relation to the ice-bed interface could increase stability. 

\section{Conclusions}

In this paper, we present a novel phase field fracture method using a Mohr-Coloumb criterion for the crack driving force, and apply this model to study cliff failure in creeping grounded glaciers.
We find that for fast-moving glaciers undergoing free slip or subject to lesser basal shear stress, damage propagates as a result of tensile longitudinal stresses in the far field regions and fractures can therefore be considered as mode I tensile crevasses. By contrast, for slow-moving glaciers subject to greater basal friction or frozen to the bedrock, fracture occurs as a result of shear stress near the terminus and may lead to full-thickness cliff failure through a mixed-mode fracture process. Different values of internal friction $\mu$ and cohesion $\tau_\mathrm{c}$ are considered in our analysis, finding that friction influences the mode of failure, whilst cohesion influences the stable cliff height. For values of $\tau_\mathrm{c} = 1$ MPa, we find that cliff failure occurs in land terminating glaciers of height $H \geq 200$ m, a result which is consistent with the surrounding literature. For $\tau_\mathrm{c} = 0.25 \; \textrm{MPa}$ cliff failure will occur if $H \geq 85$ m.

For marine-terminating glaciers, the application of ocean-water pressure results in a compressive stress that offsets extensional stresses from glacier motion, allowing for thicker glaciers to exist. Subaerial cliff failure in marine-terminating glaciers is observed, with fracture propagating to the ocean-water surface, leading to potential cliff detachment. For fractures without the presence of meltwater, cliff detachment is dependent on whether the glacier free-board exceeds a critical value. Similarly to land-terminating glaciers, the value of critical free-board is highly dependent on cohesive strength $\tau_\mathrm{c}$, but larger values of glacier free-board are observed compared to the land-terminating case. For $\tau_\mathrm{c} = 0.5 \; \textrm{MPa}$, a critical free-board of $H-h_\mathrm{w} \approx 215$ m is observed, which is in accordance with field observations for Alaskan and Greenland outlet glaciers. However, the critical glacier free-board increases to $H-h_\mathrm{w} \approx 405$ m for $\tau_\mathrm{c} = 1.0$ MPa, which may be an overly conservative estimate of glacier stability. In our future work, we will consider model intercomparison studies that investigate various cliff failure criteria to reveal the differences between continuum and discrete element approaches used in the literature.   

\subsection{Acknowledgements} 

T. Clayton acknowledges financial support from the Natural Environment Research Council (NERC) via Grantham Institute - Climate Change and the Environment (project reference 2446853). R. Duddu gratefully acknowledges the funding support provided by the National Science Foundation’s Office of Polar Programs via CAREER grant no. PLR-1847173, NASA Cryosphere award no. 80NSSC21K1003, and The Royal Society via the International Exchanges programme grant no. IES/R1/211032. T. Hageman acknowledges financial support through the research fellowship scheme of the Royal Commission for the Exhibition of 1851. E. Martínez-Pañeda acknowledges financial support from UKRI’s Future Leaders Fellowship programme [grant MR/V024124/1] 

\subsection{Author contributions}
TC: Conceptualization, data curation, formal analysis, investigation, methodology, software, visualization, writing - original draft; RD: Conceptualization, supervision, writing - review \& editing; TH: Conceptualization, methodology, supervision, writing - original draft,  writing - review \& editing; EMP: Conceptualization, funding acquisition, resources, supervision, project administration, writing - review \& editing.

\end{document}